\documentclass[12pt]{book}

\usepackage[dvips]{graphicx,color}
\usepackage{times,makeidx,subfigure,epsfig,view}

\begin{document}

\BookTitle{\itshape }
\CopyRight{}
\pagenumbering{arabic}

\chapter{The Status of VHE Astronomy}

\author{Rene~A.~Ong} 

\author{Department of Physics \& Astronomy, University of California,
Los Angeles, CA 90095, USA\\}

\AuthorContents{Rene A. Ong} 


\section*{Abstract}

This paper summarizes the status of very high-energy (VHE)
astronomy, as of early 2003.  
It concentrates on observations made by $\gamma$-ray
telescopes operating at energies above 10 GeV.
This field is an exciting one to be working in, with a growing 
and varied list of established sources that includes plerions,
supernova remnants, and active galactic nuclei.
Some of these sources are believed to be well understood
theoretically,
but others are much more puzzling.
New results include the discovery of the first unidentified 
source at very high energies
and the detection of $\gamma$-rays from 
a nearby starburst galaxy (NGC 253).
The arrival of a new generation of telescopes, both on the ground
and in space, argues that we can anticipate a wealth
of exciting results in the near future.

\section{Introduction}

This review is based on the summary talk given at a major
VHE $\gamma$-ray meeting and on the papers
presented at that meeting [1].
The review is imperfect for a number of reasons.
First, it is not comprehensive, but is rather 
a selective summary of the most interesting recent results.
Second, the review is largely confined to the experimental and 
observational picture.
For a summary of the theoretical perspective, 
see the talk by Takahara [1].

VHE $\gamma$-ray astronomy came of age in the 1990's [2].
The first solid detections of a galactic source (Crab Nebula) 
and an extragalactic 
one (Markarian 421) came in the early part of the decade.
The latter part of the decade was marked by an increasing number of 
confirmed VHE sources and an increasing variety of source type.
More importantly, it was marked by steady improvements in the quantity 
and quality of data.
The year 1997 was an exciting one in which dramatic flares from Markarian 
501 were detected
and studied
by a panoply of telescopes around the world.
By the end of the decade, detailed measurements of energy spectra and flux 
variability were routine for the stronger sources.

The 1990's also saw great progress in experimental capabilities, 
with atmospheric Cherenkov telescopes leading the way.
Arrays of multiple telescopes using higher resolution cameras 
demonstrated the
power of the Cherenkov
technique and sparked a world-wide effort to develop
the next generation of instruments.
Much of the discussion on future ground-based $\gamma$-ray 
telescopes took place
in a series of meetings entitled {\it Towards a Major Atmospheric Cherenkov Detector}.
These meetings were held
at Palaiseau (1992), Calgary (1993), Tokyo (1994), Padova (1995),
Kruger Park (1997), and Snowbird (1999).
The Kashiwa symposium is in the spirit of these past meetings, and this 
summary considers trends in the field that have taken place since 
the meeting at Snowbird.  For additional information, 
see the excellent summary from the 27th 
International Cosmic Ray Conference by Pohl [3].

\section{Big Themes}

There are a number of broad themes underlying the 
progression of the field during the last few years.
I discuss the following three themes, chosen somewhat arbitrarily:
\begin{itemize}
\item Steadily increasing source count.
\item Multi-wavelength approaches.
\item New telescopes and new instrumentation.
\end{itemize}

\subsection{VHE Source Count}

\begin{figure}[t]
  \begin{center}
    \includegraphics[height=23pc]{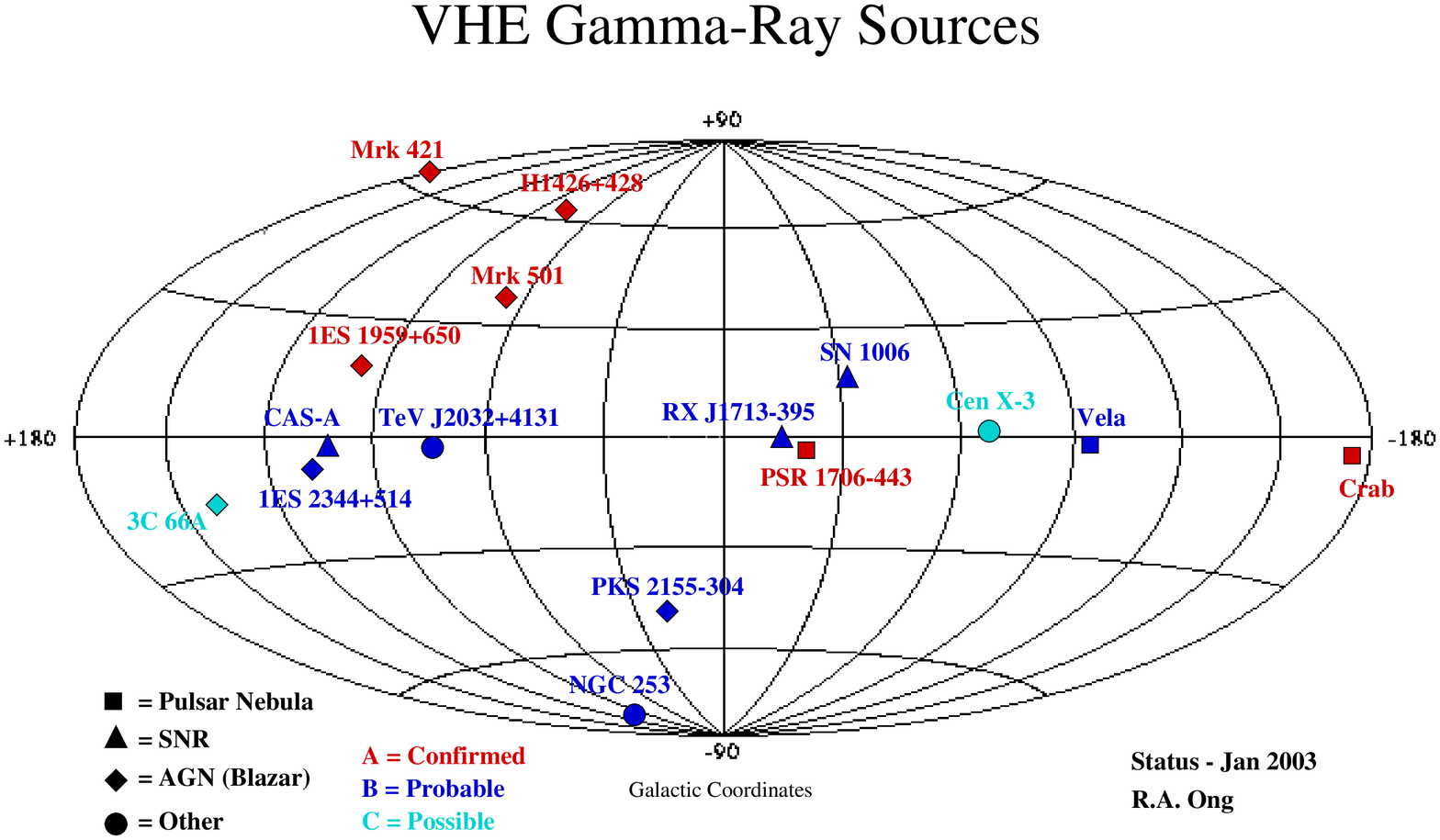}
  \end{center}
  \caption{VHE Sky Map, as of January 2003.  In the
last three years, six new sources have been reported:
the supernova remnants CAS-A and RX J1713-3946,
the blazars 1ES 1959+650 and H1426+428, 
the starburst galaxy NGC 253, 
and the unidentified source TeV J2032+4131.}
\end{figure}

The current VHE sky map is shown in Figure~1.
Since the Snowbird meeting, six new sources
have been discovered by atmospheric Cherenkov telescopes.
Of these new sources, perhaps the most exciting one is
the first unidentified VHE source - TeV J2032+4131 -
detected by the HEGRA telescope.
The other new sources are the blazars H1426+428 and 1ES 1959+650,
the supernova remnants Cassiopeia-A and RXJ 1713-3846, and
the starburst galaxy NGC 253.

Sources are detected with varying degrees of
confidence by a varying number of groups.
We can divide the sources into three categories based on
a somewhat arbitrary measure of our
confidence in their detection:
\begin{itemize}
\item {\bf Confirmed Sources (A):} Six sources have been detected by
multiple experiments at high significance levels: 
Crab, Markarian 421, Markarian 501, PSR 1706-443,
H1426+428, and 1ES 1959+650.
\item {\bf Probable Sources (B):} Eight sources have been detected 
at a high significance level by at least one group:
Vela, 1ES 2344+514, SN1006, Cassiopeia-A, RXJ 1713-3846,
TeV J2032, PKS 2155-304, and NGC 253.
\item {\bf Possible Sources (C):} Two other sources have been 
claimed: Centaurus X-3, and 3C66A.
\end{itemize}

Considering the confirmed and probable sources together, we now
have a catalog of
fourteen objects, classified in the following way:
3 pulsar nebulae, 3 supernova remnants, 1 starburst galaxy,
6 AGN, and 1 unknown.  All AGN are of the BL-Lac type, and 
all confirmed AGN detections have so far
been made in the northern hemisphere.
With the advent of the new HESS and CANGAROO-III telescopes in
the south, we can expect the southern hemisphere to become
a more important part of the overall picture.

\subsection{Multi-Wavelength Approaches}

VHE $\gamma$-ray astronomy does not exist in a vacuum and
contact with other wavebands is essential to understanding
the detailed astrophysics of the sources.
In particular, the connection to X-rays is very important
for this field because of the similar non-thermal processes
probed and because of the wealth of new data from the
current generation of X-ray satellite experiments:
RXTE, Chandra, XMM/Newton, and INTEGRAL.
Overview talks on our understanding of 
supernova remnants (SNR's) were presented
by Berezhko, Slane, and Tanimori.
They reminded us about how important it is to understand the broadband
spectral energy distribution (SED) in SNR's and plerions.
For SN 1006, the basic model of synchrotron emission from a
non-thermal electron distribution and an inverse-Compton
component appears to well describe the observed
data X-ray and $\gamma$-ray data.
By contrast, the picture for RX J1713-3846 is not clear
at all, as discussed below.

The general AGN landscape was covered by Coppi and Mukherjee.
Here, in broad terms, the X-ray and TeV $\gamma$-ray data
show correlation in their SED's and in their temporal behavior.
One is tempted to develop a unified picture to explain or
describe these correlations (see, for example, [4]).
Such schemes have merit, but things get more complicated when
one looks at these sources in detail.
For example, it now appears clear that the TeV (and GeV)
blazars have discrete flare states -- i.e. one sees flares
that last for long periods of time (weeks/months),
and then the source quiets down.
This situation is hard to explain in the context of a 
single blob model.
Among other things, we would like to understand: 1) the acceleration
energy spectrum of the primary electrons, 2)
whether the high-energy cutoff in some objects is intrinsic
or simply due to photon-photon pair-production (in the source
or in intergalactic space), and 3) the
source of the seed photons for Compton up-scattering.

Turning to the GeV sky, Pohl reminded us of the main questions
left from the EGRET mission. 
EGRET detected a significant excess in the GeV $\gamma$-ray signal
at high latitudes that is not understood.
In addition, the majority of the point sources detected by EGRET
are not identified. Probably a large fraction of these are
galactic in origin, but only minor fraction appear to be
SNR's or pulsars.  Thus it is likely that a new 
high-energy galactic source class exists.
Are these two problems (the GeV excess and the unknown nature of
most EGRET sources) related?
We do not know the answer to this question. 
TeV observations can shed light 
on this puzzle, even before the launch of GLAST. 

\begin{figure}[t]
  \begin{center}
    \includegraphics[height=25pc]{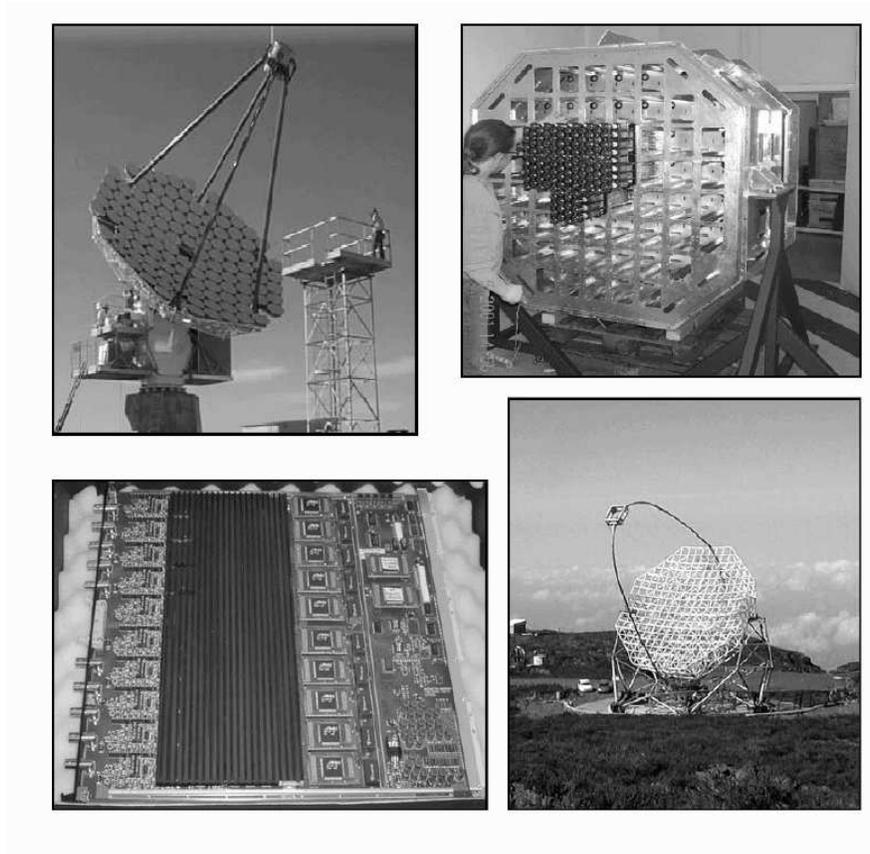}
  \end{center}
  \caption{New Cherenkov telescope instrumentation.
Upper left: second telescope of the CANGAROO-III
array. 
Upper right: camera integration for one of the HESS telescopes.
Lower left: readout electronics for VERITAS (500 MHz FADC).
Lower right: large 17m diameter telescope of MAGIC.}
\end{figure}

\subsection{New Telescopes and Instrumentation}

There were many interesting talks on new
detector techniques at this meeting. 
Most talks were related to the atmospheric
Cherenkov technique, and they described
details of the instrumentation for
individual telescopes or for new detector concepts.
The most important development is the
great progress made on the construction of the 
next generation of telescopes:
CANGAROO (Ohishi et al.), 
HESS (Hofmann et al.), 
MAGIC (Lorenz et al.), and
VERITAS (Ong et al.).
Some images of the instrumentation being built for these
four telescopes are shown in Figure~2.

The {\bf CANGAROO-III} array is being constructed in
Woomera, Australia (31.1$^\circ$S, 137.8$^\circ$E, 160$\,$m a.s.l.).
The array will comprise four $10\,$m diameter telescopes, each having
an imaging camera of $430$ pixels covering a field of
view of 4$^\circ$.
The optical reflectors are parabolic in shape, with mirror
facets made of glass fiber reinforced plastic.
The camera consists of conventional photomultiplier tubes (PMT's)
with each tube covering a full angle of 0.17$^\circ$ on the sky.
The electronics are largely conventional VME-based modules.
Two telescopes of the array are already in place at Woomera and
are being commissioned.
The construction of the third and fourth telescopes is scheduled
for 2003.
Observations with all four telescopes are expected
to start in 2004.

The {\bf HESS} array is well underway at a site in the Gamsberg 
area of Namibia (23.3$^\circ$S, 16.5$^\circ$, 1800$\,$m a.s.l.).
HESS will consist of four 12$\,$m diameter
telescopes, each with an imaging camera of 960 pixels covering
a 5$^\circ$ field of view.
The optical reflectors use the Davies-Cotton design with a focal
length of $15\,$m.
The mirror segments are made of ground glass that are then
aluminized and quartz coated.
The camera is a fully integrated and modular assembly, comprising
PMT's and electronics (CPU, trigger, and readout) in the camera
body.
The readout is accomplished by the Analog Ring Sampler (ARS)
device which samples the PMT signal at high rate to provide
a measure of the charge integrated over a 16$\,$ns window.
HESS presented initial data from two complete telescopes.
The construction of the third and fourth telescopes is scheduled for
2003, with four-telescope observations starting in 2004.

The {\bf MAGIC} telescope is under construction on
the island of La Palma (28.8$^\circ$N, 17.8$^\circ$W, 
2225$\,$m a.s.l.).
The telescope comprises a number of advanced technologies
in the design of the mirror, the camera, and the readout
electronics.
The large 17$\,$m diameter reflector has a parabolic profile,
and the individual mirror segments
are made from diamond-turned aluminum blanks.
The camera uses 574 PMT's covering a 4$^\circ$ field
of view.
Camera readout is accomplished by signal transmission through
analog optical fibers to a 300$\,$MHz Flash-ADC (FADC) system.
Much of MAGIC is already in place, and completion of the
full mirror and camera are set for spring 2003, with first observations
to follow later in 2003.

{\bf VERITAS} is an array of telescopes to be constructed
near the base camp of Mt. Hopkins in Arizona, USA 
(31.7$^\circ$N, 110.9$^\circ$W, 1300$\,$m a.s.l.).
The first phase of the project, VERITAS-4, will consist of
four 12$\,$m diameter telescopes, each with a 499 pixel
camera covering a field of view of 3.5$^\circ$.
In the second phase, three more telescopes will be
added.
The optical reflectors use the Davies-Cotton design
with a focal length of $12\,$m.
Mirror facets are made of ground glass that has been
aluminized and anodized.
PMT signals are amplified in the base of the phototube
and transmitted via high-bandwidth cable to
a system of custom-built 500 MHz FADC's.
Construction on the first VERITAS telescope is underway,
with initial operation scheduled for late 2003.
Current plans call for a four-telescope array to be operational
in late 2005.

In space, there are a number of existing or planned
instruments that can
probe the high-energy gamma-ray regime.
Chief among these is {\bf GLAST} [5], scheduled for
launch in the fall of 2006.
GLAST will have a substantially larger collection area,
a wider field of view, and a smaller dead time
than EGRET on the
Compton Gamma Ray Observatory.
In comparison with EGRET,
GLAST will also have superior angular resolution and a
wider spectral range (up to 100 GeV and beyond).
Other planned instruments include AGILE, AMS, and SWIFT.

\section{New Results}

There have been many nice results reported in the last
year.  I have selected a number of them to discuss in this
summary:

\begin{enumerate}
\item TeV J2032+4131: the first unidentified TeV $\gamma$-ray source,
\item the mystery of SNR RX J1713-3946,
\item VHE sky surveys,
\item new AGN results,
\item NGC 253: a starburst galaxy source of VHE emission, and
\item spectral measurements between 50 and 250 GeV.
\end{enumerate}

\subsection{The First Unidentified TeV Source: TeV J2032+4131}

\begin{figure}[t]
\subfigure[{\protect\bf Fig. 3a.}  
Detection of unidentified source
TeV J2032+4131 by HEGRA.
The spectrum of the source is shown, along with
the spectrum from the nearby
EGRET source 3EG J2033+4118.
Also shown are
upper limits from ASCA-GIS and HEGRA-AIROBICC (H-A).
]
{ \epsfig{file=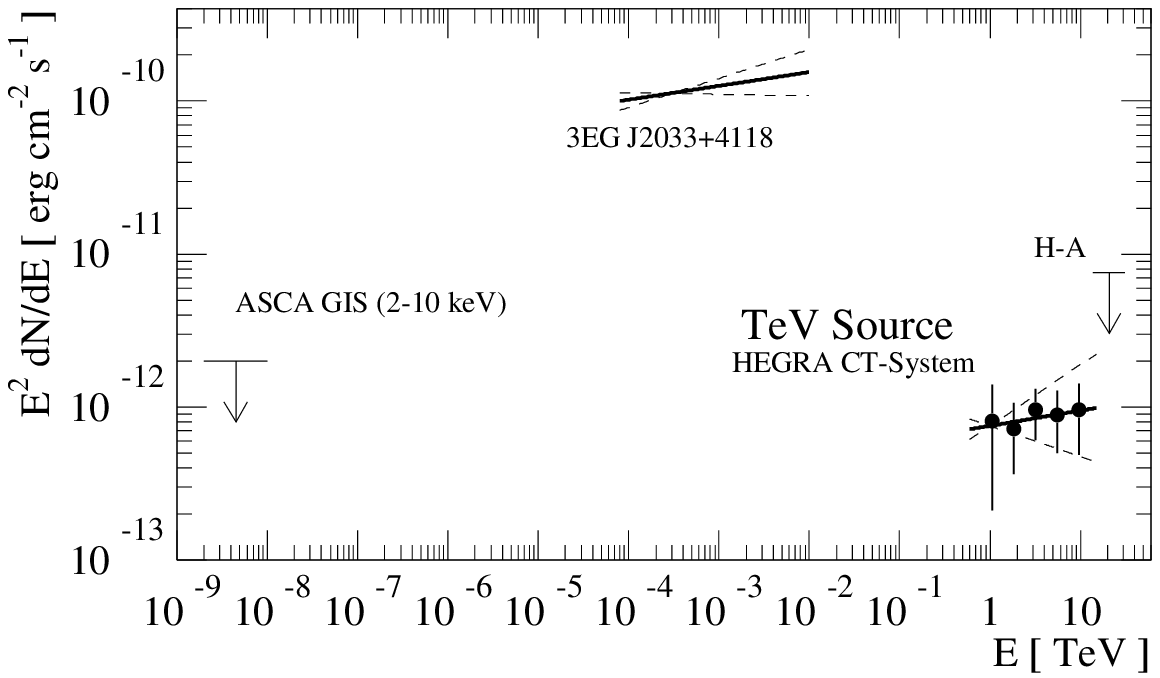,width=3.0in}}
\subfigure[{\protect\bf Fig. 3b.} 
Map of the TeV $\gamma-$ray 
significance (color shades) for the region around 
TeV J2032+4131, along with the
2$\sigma$ contour for the source.
Also shown are the 95\% error ellipses for various EGRET 
sources, the core of the Cygnus OB2 complex, 
and an overlay of ASCA GIS X-ray data (contours).
]
{ \epsfig{file=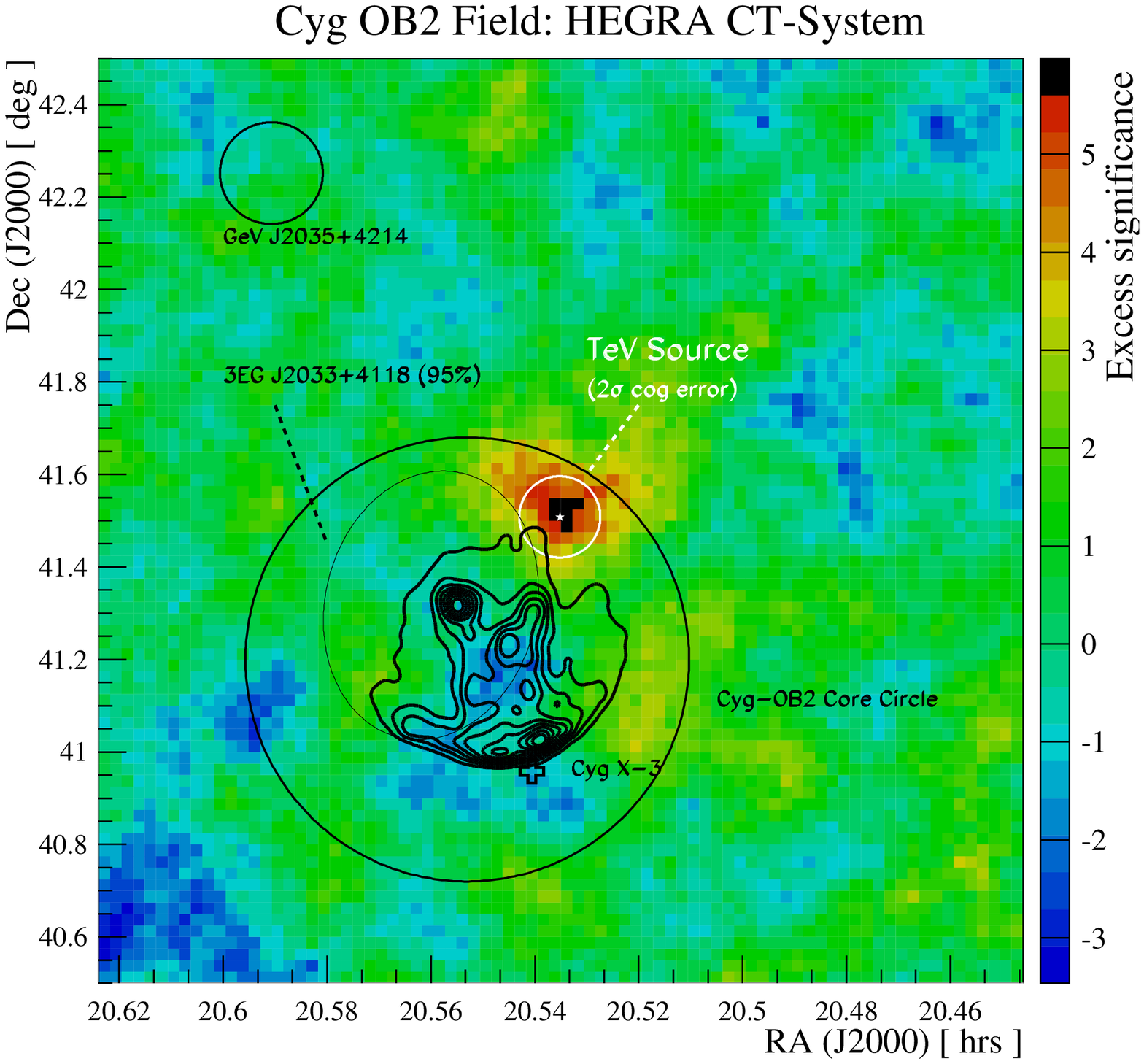,width=2.8in}}
\end{figure}

HEGRA presented evidence for the first unidentified 
$\gamma$-ray
source at TeV energies (Rowell, see also [6]).
The source was spotted in scans of the region near the famous Cygnus X-3.
After accounting for trials factors, the detection has a statistical
significance of 4.6$\sigma$.  The source is weak, but it appears to be 
steady with an
integral flux of $\sim 30\,$mCrab.

This new source has several interesting properties.
Chief among them is that it is not clearly identified with any EGRET source
or any known X-ray source.  As shown in Figure~3, the source is relatively
far away from Cygnus X-3, but it is close to the Cygnus OB2 complex
and at the edge of the error box of the EGRET source 3EG J2033+4118.
A second interesting feature of the source is its hard spectrum
with a differential index $\alpha \sim -2$.
A third feature is that the source appears to be somewhat
extended (approximately 6 arc-minutes across). Newer data
from HEGRA taken in 2002 appear to confirm the source detection and
spectrum.

There are a number of factors that argue for a galactic origin for 
TeV J2032+4131, but the basic fact is that we just cannot
yet make a firm identification.
There is the exciting possibility that this source is a new type of
object not seen at other wavelengths.
In any case, the catalog of unidentified TeV sources has been started.

\subsection{The Mystery of RX J1713-3946}

The CANGAROO group reported the detection of VHE $\gamma-$rays
from the supernova remnant RX J1713-3946 (SNR G347-0.5) 
from data taken with the original  
3.8$\,$m telescope and with the newer 7$\,$m telescope
(Kawachi, see also [7]).
The group argues that the observed TeV spectrum 
is not consistent with that expected from
inverse-Compton scattering from accelerated electrons, but rather is
consistent with that expected from $\pi^\circ$-decay (where the
$\pi^\circ$'s are produced when accelerated protons collide with ambient
material).
Thus, the detection of this SNR can be seen as evidence for
the acceleration of very high-energy cosmic rays.

Unfortunately, the interpretation of this source may not be straightforward.
Reimer and Pohl present the case that the pion-decay spectrum that
fits the TeV data is inconsistent with the EGRET data at
GeV energies [8].
Even though EGRET did not detect this object, the EGRET sky map
in its vicinity (in particular, near the unidentified
source 3EG J1714-3857)
provides an upper limit on the GeV flux at the source location.
This upper limit appears to be below the expected flux from the
pion-decay model (see also [9]).

At this meeting, Slane showed that inverse-Compton models could
fit the GeV and TeV data if one assumed a relatively
large value for the magnetic field (B $\sim 50\,\mu$G) with
a relatively small filling factor.
This might be a reasonable picture for a supernova remnant that
is evolving towards a molecular cloud.

In the mind of this reviewer, the origin for
the TeV radiation from this source has not been clearly established.
In general, supernova remnants will have both an electron inverse-Compton
component and a proton pion-decay component.
Separating and identifying
these two components is challenging.
Clearly, the new, more powerful Cherenkov telescopes in the southern
hemisphere (HESS and CANGAROO-III) will soon be able to shed light
on the question of cosmic-ray acceleration in SNR's.

\subsection{VHE Sky Surveys}

\setcounter{figure}{3}
\begin{figure}[t]
  \begin{center}
    \includegraphics[height=15pc]{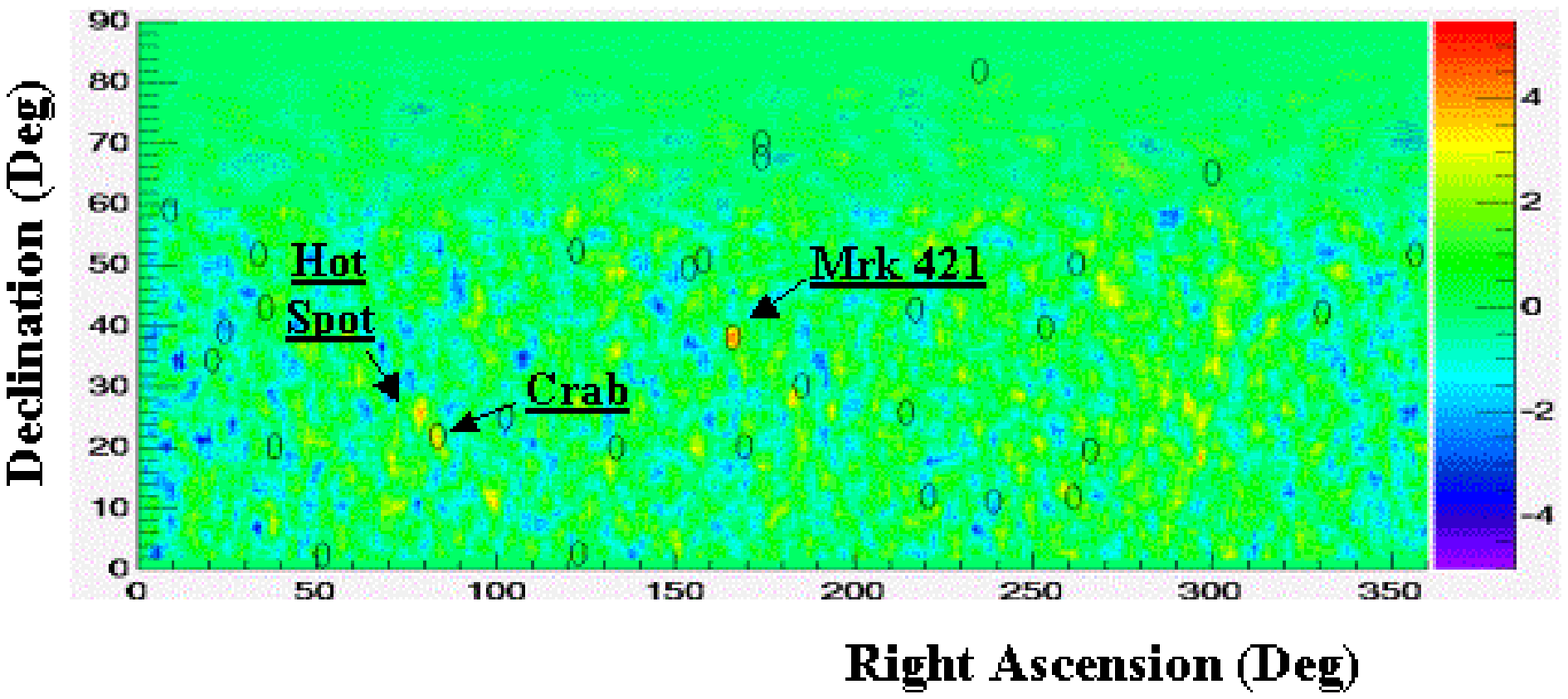}
  \end{center}
  \caption{Map of the northern hemisphere VHE sky
as seen by Milagro.  The excess significance (color
scale on right hand side) is plotted
as a function of equatorial coordinates.
The brightest locations are Mrk 421, the Crab, and
an unidentified location (``Hot Spot'').
The circles mark locations of AGN selected in [10].
The threshold energy of the experiment is
$\sim 4\,$TeV in the broad declination range between
20$^\circ$-50$^\circ$; it increases outside this range.}
\end{figure}

Our knowledge of the VHE sky is still very incomplete.
Atmospheric Cherenkov telescopes have achieved excellent
point source sensitivity, but have so far surveyed only
a small fraction of the overhead sky.
This fraction is steadily increasing, and HEGRA
presented preliminary results from analysis
of archival data covering 0.4 sr (3.5\% of the sky).
In this study, two different approaches to flat-fielded
the data by using contemporaneous background observations were
used.
The two approaches generally gave similar results.
The HEGRA data showed a handful of sky locations
having a statistical significance in the excess event count 
greater than four standard deviations (Puehlhofer).
The strongest significances can be identified with known objects
(Cassiopeia-A, H1426+428, and TeV J2032+4131), and
there is no evidence for any new strong sources in this
search.

Much larger fractions of the sky can be surveyed by
air shower detectors.
These detectors have wide fields of view and high
duty-cycle operation, but weaker sensitivity
relative to Cherenkov telescopes.
New results were presented from the Tibet air shower array
(Sakata) and from the Milagro experiment (Sinnis).
Both experiments have surveyed the northern hemisphere sky
at $\gamma$-ray threshold energies above 3-4$\,$TeV.
The results from the two experiments are similar, namely
that there are few, if any, sources that have a time-averaged
VHE $\gamma$-ray flux brighter than the Crab Nebula.
The Milagro sky map for one year of data is shown in Figure~4.
It would be valuable to compare the sky maps reported by the
Tibet and Milagro experiments to search for possible
correlations -
the results from such a comparison were not reported.

\subsection{New AGN Results}

\begin{figure}[t]
\subfigure[{\protect\bf Fig. 5a.}  
Measurement of the spectrum of the BL-Lac object
H1426+428 at very high energies by VERITAS.
Various fits to the data are shown, as indicated in 
the legend.
]
{ \epsfig{file=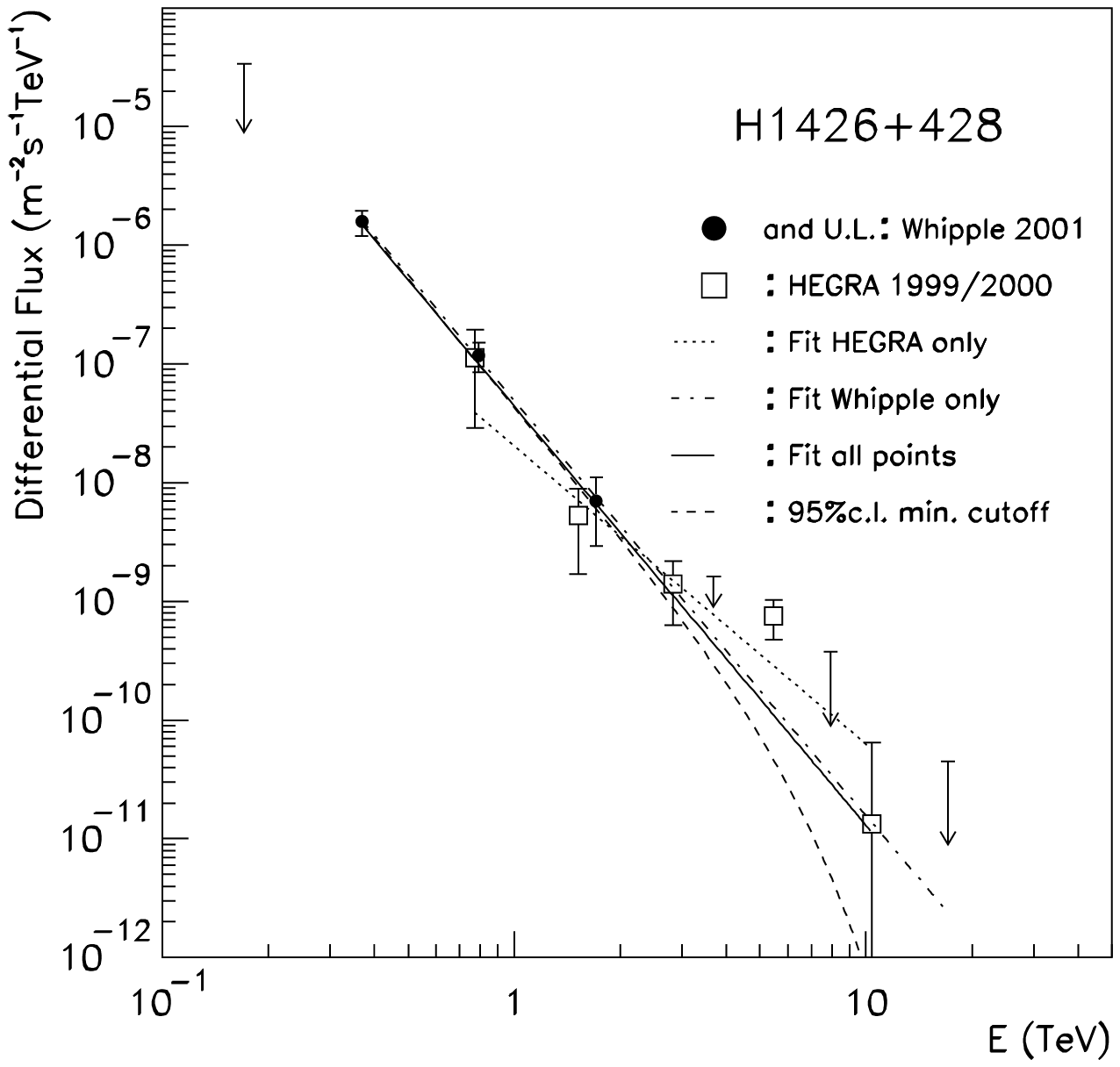,width=2.7in}}
\subfigure[{\protect\bf Fig. 5b.} 
The spectrum of H1426+428 measured by HEGRA.  The
curves correspond to assumed spectra before and after the effect
of absorption by the extragalactic background light (EBL),
for two different EBL spectral energy distributions.
]
{ \epsfig{file=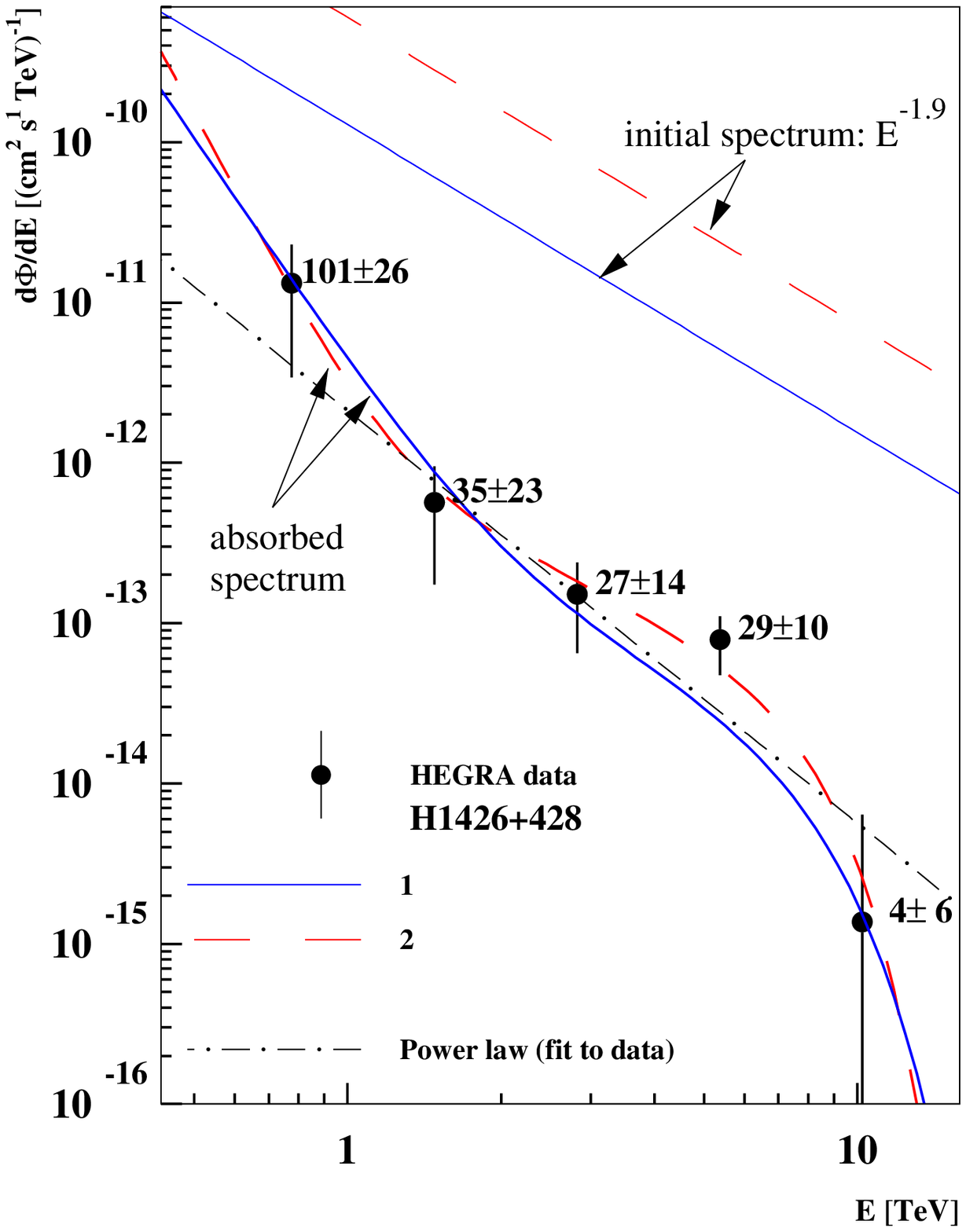,width=2.6in}}
\end{figure}

In the last two years, two new AGN have been detected in VHE $\gamma-$rays.
The first of these, H1426+428, is an X-ray selected BL-Lac object that is
the most distant source yet detected at very high energies ($z=0.129$).
The source was monitored over several years because of the high synchrotron
peak frequency (E$ > 100$ keV) seen in the X-rays, and detections were
reported by the VERITAS 
(Krennrich, see also [11,14]), 
HEGRA (Horns, see also [12]), 
and CAT [13] collaborations.
H1426+428 is weak at TeV energies and measurements 
indicate a very soft spectrum (differential index $\alpha \sim -3.0$),
as shown in Figure~5.
The spectral shape generally supports the idea that the VHE 
$\gamma$-ray signal
is heavily attenuated from absorption by the extragalactic background light (EBL).
The HEGRA group has attempted to constrain the EBL spectral energy distribution
(SED) by starting with an intrinsic source spectrum and fitting to the observed
spectrum.  This technique yields a reasonable shape
and normalization for the EBL SED.

The other newly detected AGN is 1ES 1959+650.
This source is an X-ray selected BL-Lac object at a redshift $z = 0.048$.
It is a weak EGRET source that exhibits a high degree of variability in
X-rays.
Early evidence for this source was
reported by the Telescope Array [15] and by HEGRA [16]. 
In May 2002, the source went into a flaring state
and strong VHE $\gamma-$ray emission was reported by HEGRA.
Initial results
on the spectra and light curve for this source were
presented by HEGRA (Horns) and VERITAS (Holder)
at this meeting.
VERITAS sees a spectral index similar to that for the Crab and strong
night-to-night variability [17].
On one night, a doubling time scale of $\sim 7$ hr was observed.
HEGRA sees a softer spectral index relative to the Crab, strong
variability, and evidence for spectral hardening.
The last result is very interesting; in the low state, the source
spectra appears to be well fit by a power law of differential
index $\alpha \sim -3.3$, but in the high state, a simple power law
is a poor fit to the data and a curved spectral shape is required.
Analyses of multi-wavelength observations of 1ES 1959+650 are in progress.

In addition to detecting new AGN at very high energies, we are continuing
to learn more about the now well-known TeV sources Markarian 421 and
Markarian 501.
Mrk 421 exhibited dramatic flaring activity in 2001, and the
large flux variations permitted detailed spectral measurements
to be made as a function of flux level.
These measurements clearly show spectral variability,
where the spectrum hardens as the flux level increases.
This behavior could be expected in a model where the synchrotron
peak gets shifted to higher energy during the flaring activity.
Similar results on the spectral variability of Mrk 421
were presented by VERITAS (Krennrich) and HEGRA (Horns).
As shown in Figure~6, the differential spectral index varies
from $\alpha \sim -3.0$ at relatively low flux values to
$\alpha \sim -2.0$ at the highest recorded flux levels.
Spectral variability is also observed on short time scales
($\sim 0.5\,$hr),
but there is a lot of scatter in the
data.
HEGRA also reported a possible diurnal spectral variability
indicating some sort of hysteresis-like effect.
We are reminded that these blazar systems are quite complicated and
that more data of this quality are needed to fully understand
these sources.

\begin{figure}[t]
\subfigure[{\protect\bf Fig. 6a.}  
Spectral variability of Markarian 421, as measured
by the HEGRA CT System.  The spectral index is plotted
as a function of flux level for various epochs.
]
{ \epsfig{file=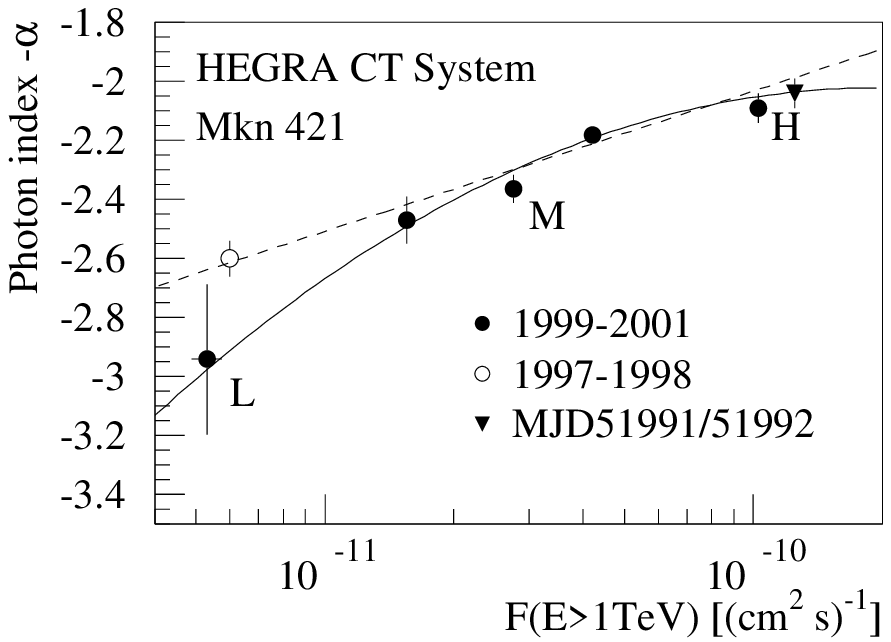,width=3.0in}}
\subfigure[{\protect\bf Fig. 6b.} 
Spectral variability of Markarian 421, as measured by
VERITAS.  The spectral index is plotted as a function 
of the flux level in units of the Crab flux.
]
{ \epsfig{file=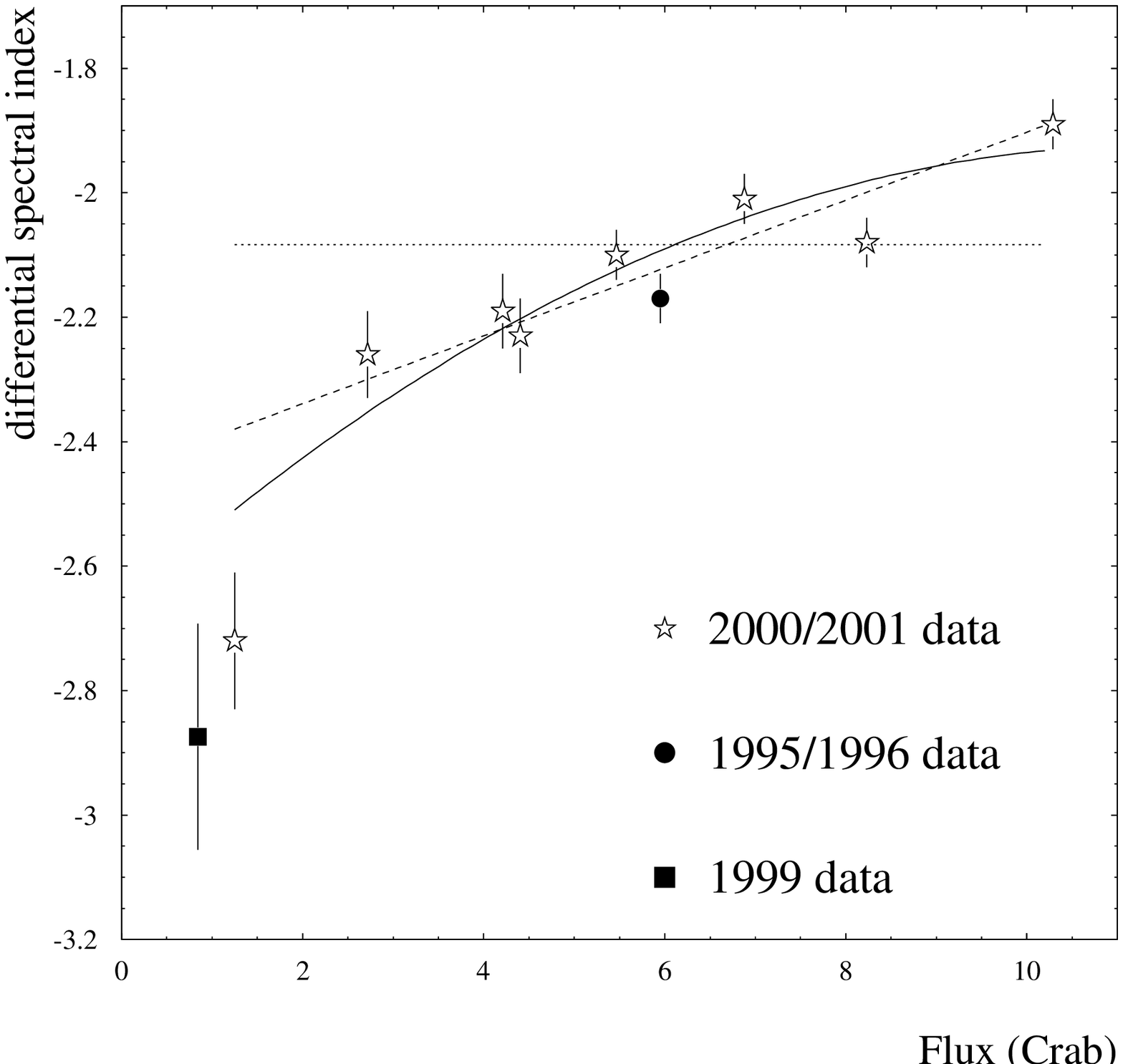,width=2.8in}}
\end{figure}

\subsection{New Source: NGC 253}

\begin{figure}[t]
\subfigure[{\protect\bf Fig. 7a.}  
CANGAROO observations of the starburst galaxy NGC 253.
The image orientation parameter, $\alpha$, is shown
for loose (left) and tight (right) selection cuts.
The points (shaded histogram) correspond to the
on (off) source regions.
]
{ \epsfig{file=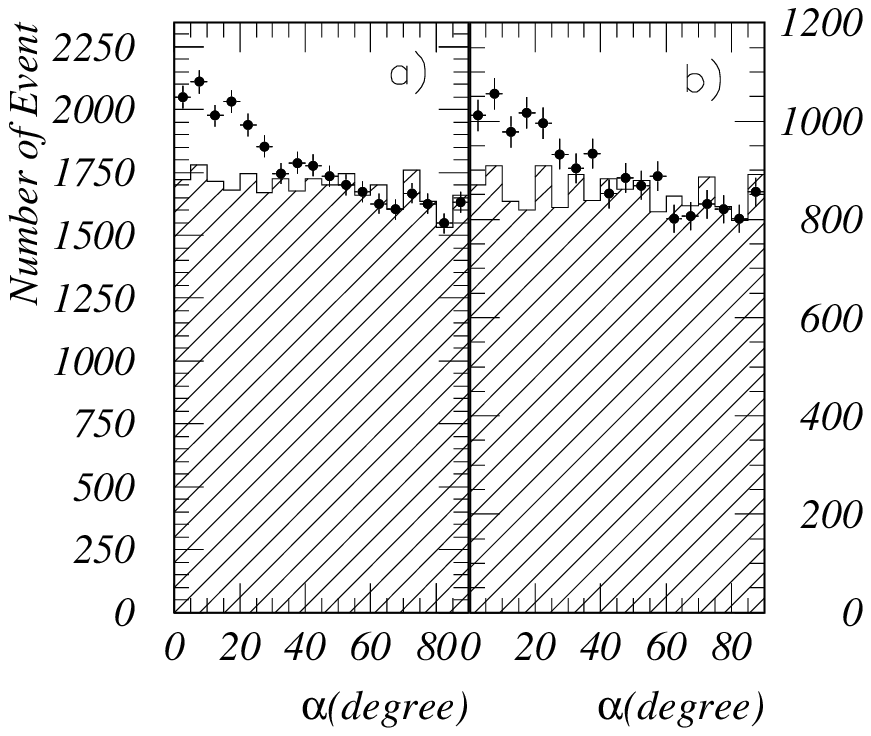,width=2.6in}}
\subfigure[{\protect\bf Fig. 7b.} 
Multi-waveband observations of NGC 253 taken at different
epochs.  The CANGAROO data are shown by the squares.
Observations at other wavebands are identified
in the legend.
]
{ \epsfig{file=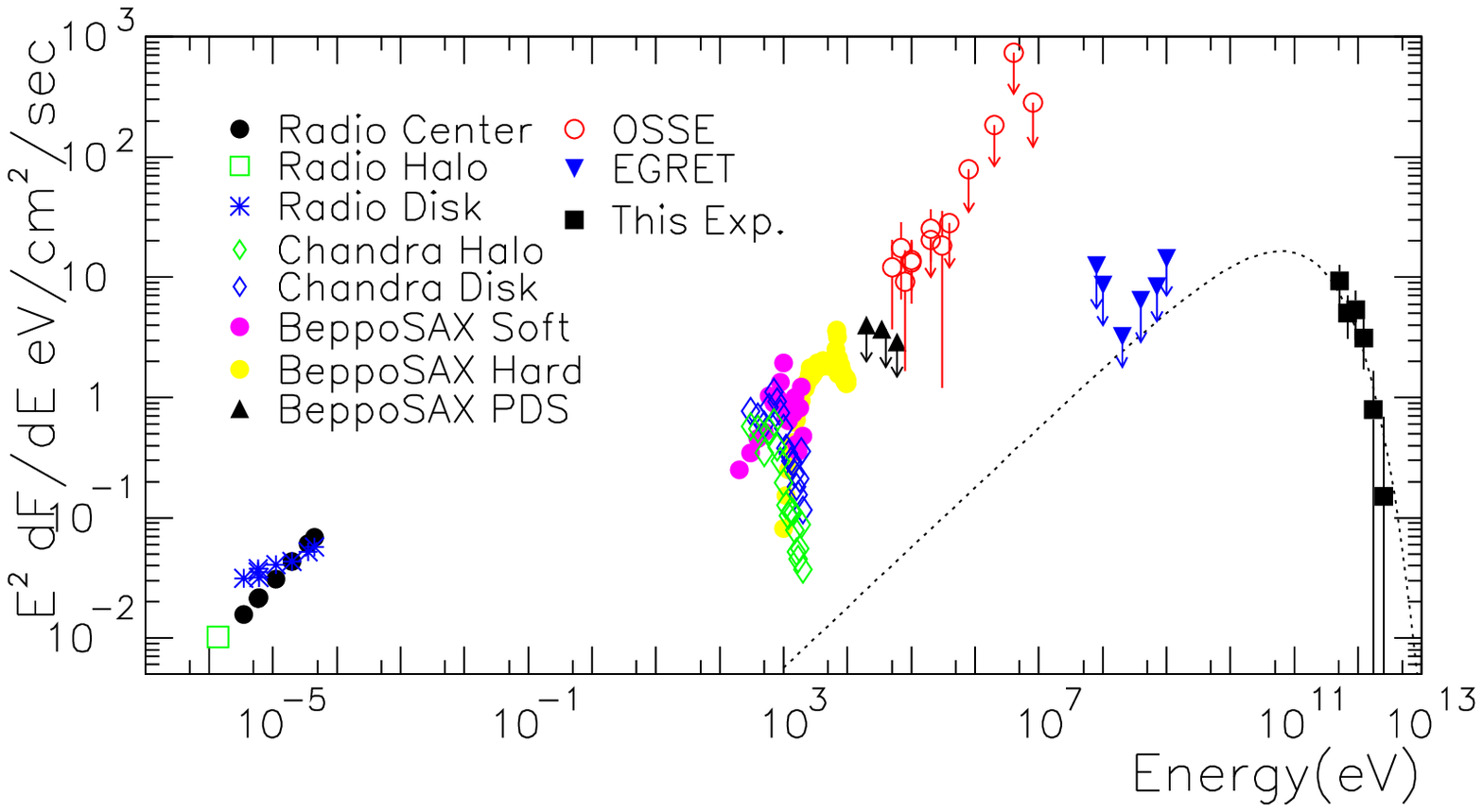,width=3.2in}}
\end{figure}

At this meeting, the CANGAROO collaboration reported the
detection of VHE $\gamma-$rays from the starburst galaxy NGC 253 
(Itoh, see also [18]).
The group also presented evidence for possible
$\gamma-$ray emission from the galactic
center (Tsuchiya)
and from two supernova remnants:
RCW86 (Watanabe)
and RX J0852-4622 (Katagiri).

At a distance of $\sim 2.5\,$Mpc,
NGC 253 is one of the closest starburst galaxies.
The galaxy is known to have a very high
star formation rate, at the level of 0.1-0.3 SN/year.
Although EGRET did not detect this source at GeV
$\gamma-$ray energies, a large and bright synchrotron halo is 
seen in the radio band.
Recent Chandra data from the core region indicates a heavily absorbed
source of hard X-rays that
may possibly signal a weak active nucleus [19].
It is argued (Voelk)
that the high star formation rate in NGC 253 will
lead to an enhanced rate of cosmic ray production (factor of
10-100 times greater than the Milky Way) and thus a substantial
level of diffuse $\gamma-$ray emission.

CANGAROO made observations of NGC 253 totalling $\sim 75\,$hr over a
two year period.  The distribution of the image orientation 
parameter $\alpha$ is shown for these data in Figure~7.
An excess of events is observed at low values of $\alpha$, corresponding
to a raw significance of $11.1\sigma$ and a detection 
rate of 0.56 $\gamma-$rays/min.  The source appears to be spatially extended in comparison
with the Crab control data.  
A simple power-law fit to the energy spectrum does not adequately
describe the data, and a strong turn-over below 1 TeV is needed.
The multi-waveband spectrum of NGC 253, shown
in Figure~7, indicates that the hard X-ray and TeV $\gamma-$ray data
cannot be described by a simple model of inverse-Compton emission
from a single source of electrons.
The TeV $\gamma-$rays may possibly originate from the extended halo and
the X-rays from the central region.
Further observations of this source will be important to understanding
its high-energy cosmic ray environment.

\subsection{Spectral Measurements Between 50 and 250 GeV}

The atmospheric Cherenkov technique has been a powerful technique
for detecting sources and measuring their spectra at energies
above several hundred GeV.
Very recently, this technique has been successfully applied by
solar array telescopes
at energies between 50 to 250 GeV.
These telescopes give us the first source measurements in this unexplored 
waveband.
Both CELESTE and STACEE presented results from recent
observations.

CELESTE reported results from a spectral analysis of 
Crab Nebula and Markarian 421 data (Piron).
An energy resolution of $\sim 20$\% was achieved.
The CELESTE data provide spectral measurements near the peak of
the putative inverse-Compton component, and the measured spectra are
observed to connect well with higher energy results from the CAT
telescope.
Future work is needed to reduce the systematic uncertainties
due to varying atmospheric conditions.
CELESTE has also made low-threshold observations to search for the
Crab Pulsar.
Although the pulsar has not yet been detected by a ground-based
$\gamma-$ray telescope, CELESTE is the first to explore
the very interesting region below 50 GeV.

STACEE reported results from detailed observations made on 
Markarian 421 during the 2001 flare 
(Hanna, see also [20]).
The light curve showed
night-to-night flux variations and peak detection rates
exceeded 1000 $\gamma-$rays/hr.
The average flux measured by STACEE generally agrees with an extrapolation
from the VERITAS spectrum recorded on the same nights.
Moreover, a good correlation is observed between the STACEE and 
VERITAS flux measurements on a daily
time scale.  
Weaker correlation is seen between the STACEE and RXTE light curves.
STACEE also reported on observations made of the AGN sources
W Comae ($z = 0.102$) and H1426+428.
Neither of these sources were detected, but the limits are interesting and
constrain the spectral behavior in the waveband centered on 100 GeV.

\section{Summary}
VHE astronomy continues to be an exciting area to work in.
The catalog of known sources is growing, and our knowledge
of specific source properties has significantly increased 
over the
last few years.
We are discovering new types of sources and measuring their
spectral, temporal, and multi-wavelength behavior with greater
accuracy.
With the advent of a new generation of ground-based Cherenkov telescopes,
we are entering a new era.
VHE astronomy is observationally driven, and
the new telescopes will soon be providing lots of results to be
understood by the theorists.

It is very gratifying to see the 
great progress made on the
construction of the CANGAROO, HESS, MAGIC, and
VERITAS telescopes.
It has taken ten years to reach the ``Major
Atmospheric Cherenkov Detector'' but we are almost there!
Of course, nothing stands still, and work is already well underway
towards a future generation of instruments, possibly Cherenkov
telescopes at high altitude with very wide fields of view.
New technology in optics, photodetectors, and electronics
will be essential for moving the field forward.

\section{Acknowledgements}
I wish to acknowledge organizers of this symposium,
Ryoji Enomoto, Masaki Mori, and Shohei Yanagita,
for inviting me to give this review.
Many thanks are due to all the speakers who gave me materials before,
during, and after the meeting,
especially Ryoji Enomoto,
Dieter Horns, Frank Krennrich, Gavin Rowell, and Gus Sinnis.
Dirk Petry kindly provided me with a template for the sky map.
Any inaccuracies in the presentation of results are my responsibility.

Considerable gratitude is owed to Tadashi Kifune, one
of the pioneers in the field of
using ground-based telescopes to study VHE
astrophysics.  Tadashi has been a key figure in the exploration
of the southern hemisphere sky.
With the advent of new telescope arrays,
we can rightfully expect to see an explosion in the number of
VHE $\gamma-$ray sources, as predicted by Tadashi.

\section{References}

\re 1.\ 
Symposium entitled:
{\it The Universe Viewed in Gamma-Rays},
25-28 Sep 2003, Kashiwa, Japan, published by
Universal Academy Press, Inc. (Tokyo), ed. R. Enomoto.
References to papers presented at this meeting 
are given in the body of the text by listing the presenter's name.

\re 2.\ For reviews of the field, see, for example:
R.A. Ong 1998, Phys. Rep. 305, 93;
Weekes, T.C. 2000, Proc. High Energy Gamma-Ray Astronomy 
(Heidelberg),
American Institute of Physics Conf. Proc. 558
(ed. F.A. Aharonian and H.J. Voelk), p. 15.

\re 3.\ Pohl, M. 2001, Proc. 27th Int. Cosmic Ray Conf. (Hamburg),
Copernicus Gesellschaft, Invited and Rapporteur Talks, astro-ph/0111552.

\re 4.\ Donato, D., Ghisellini, G., Tagliaferri, G., and
Fossati, G. 2001, A\&A 375, 739.

\re 5.\ Kniffen, D.A. et al. 2000,
Proc. of GeV-TeV Gamma-Ray Astrophysics Workshop,
Towards a Major Atmospheric Cherenkov Detector VI (Snowbird),
American Institute of Physics Conf. Proc. 515, 
(ed. B.L. Dingus, M.H. Salamon, and D.B. Kieda),
p. 492.

\re 6.\ Aharonian, F. et al. 2002, A\&A 393, L37.

\re 7.\ Enomoto, R. et al. 2002, Nature 416, 823.

\re 8.\ Reimer, O. and Pohl, M. 2002, A\&A 390, L43.

\re 9.\ Butt, Y.M. et al. 2002, Nature 418, 499.

\re 10.\ Costamante, L. and Ghisellini, G. 2002, A\&A 384, 56.

\re 11.\ Horan, D. et al. 2002, ApJ 517, 753.

\re 12.\ Aharonian, F. et al. 2002, A\&A 384, L23.

\re 13.\ Djannati-Atai, A. et al. 2002, A\&A 391, L25.

\re 14.\ Petry, D. et al. 2002, ApJ 580, 104.

\re 15.\ Nishiyama, T. et al. 1999, Proc. 26th Int. Cosmic Ray
Conf. (Salt Lake City) 3, p. 370.

\re 16.\ Konopelko, A. 2002, American Physical Society (APS) April Meeting and
American Astrophysical Society (AAS) HEAD Meeting,
Albuquerque, NM, 20-23 Apr 2002 (unpublished).

\re 17.\ Holder, J. et al. 2003, ApJ 583, L9.

\re 18.\ Itoh, C. et al. 2002, A\&A 396, L1.

\re 19.\ Weaver, K.A. et al. 2002, ApJ 576, L19.

\re 20.\ Boone, L. et al. 2002, ApJ 579, L5.


\endofpaper
\end{document}